\documentclass[12pt]{article}

\usepackage{tabulary,graphicx,amsmath,amsfonts,amssymb}
\usepackage[utf8x]{inputenc}
\usepackage[T1]{fontenc}
\usepackage{siunitx}
\usepackage{cite}
\usepackage{amsmath}
\usepackage{subfigure}
\usepackage{caption}
\usepackage{graphicx}
\usepackage{bm}
\usepackage{float}
\usepackage{url,multirow,morefloats,floatflt,cancel,tfrupee}
\makeatletter
\usepackage{multicol}
\usepackage{colortbl}
\usepackage{xcolor}
\usepackage{pifont}
\usepackage[nointegrals]{wasysym}
\usepackage{abstract}
 
\title{\textbf{\Large 113 km absolute ranging with nanometer precision}}
\date{}

\begin{document}

\maketitle
\setlength{\parindent}{0pt}
\author
{
	Yan-Wei Chen\textsuperscript{1,2,3,*}, Meng-Zhe Lian\textsuperscript{1,2,3,*}, Jin-Jian Han\textsuperscript{1,2,3}, Ting Zeng\textsuperscript{1,2,3}, Min Li\textsuperscript{1,2,3}, Guo-Dong Wei\textsuperscript{1,2,3}, Yong Wang\textsuperscript{4}, Yi Sheng\textsuperscript{1,2,3}, Ali Esamdin\textsuperscript{4}, Lei Hou\textsuperscript{1,2,3}, Qi Shen\textsuperscript{1,2,3}, Jian-Yu Guan\textsuperscript{1,2,3}, Jian-Jun Jia\textsuperscript{3,5}, Ji-Gang Ren\textsuperscript{1,2,3}, Cheng-Zhi Peng\textsuperscript{1,2,3}, Qiang Zhang\textsuperscript{1,2,3}, Hai-Feng Jiang\textsuperscript{1,2,3}, Jian-Wei Pan\textsuperscript{1,2,3}\\
	\\
	\normalsize{$^{1}$Hefei National Research Center for Physical Sciences at the Microscale and School of Physical Sciences, University of Science and Technology of China, Hefei 230026, China }\\
	\\
	\normalsize{$^{2}$Shanghai Research Center for Quantum Science and CAS Center for Excellence in Quantum Information and Quantum Physics,University of Science and Technology of China, Shanghai 201315, China}\\
	\\
    \normalsize{$^{3}$Hefei National Laboratory, University of Science and Technology of China, Hefei 230088, China}\\
    \\
    \normalsize{$^{4}$Xinjiang Astronomical Observatory, Chinese Academy of Sciences, Urumqi 830011, China}\\
     \\
    \normalsize{$^{5}$Key Laboratory of Space Active Opto-Electronic Technology, Shanghai Institute of Technical Physics, Chinese Academy of Sciences, Shanghai 200083, China}\\
     \\
     \normalsize{$^\ast$These authors contributed equally to this work.}
}

\begin{abstract}
Accurate long-distance ranging is crucial for diverse applications, including satellite formation flying, very-long-baseline interferometry, gravitational-wave observatory, geographical research, etc~\cite{white2000imaging, cash2000laboratory, gendreau2003maxim, lawson2007terrestrial, turyshev2009laser, akiyama2019first, gomez2022probing, abbott2016observation, pritchard2002satellite, flechtner2021satellite, reigber2002high, tapley2004grace}. The integration of the time-of-flight mesurement with phase interference in dual-comb method enables high-precision ranging with a rapid update rate and an extended ambiguity range. Pioneering experiments have demonstrated unprecedented precision in ranging, achieving 5 nm @ 60 ms for 1.1 m and 200 nm @ 0.5 s for 25 m~\cite{coddington2009rapid, trocha2018ultrafast, suh2018soliton}. However, long-distance ranging remains technically challenging due to high transmission loss and noise. In this letter, we propose a two-way dual-comb ranging (TWDCR) approach that enables successful ranging over a distance of 113 kilometers. We employ air dispersion analysis and synthetic repetition rate technique to extend the ambiguity range of the inherently noisy channel beyond 100 km. The achieved ranging precision is 11.5 $\mu$m @ 1.3 ms, 681 nm @ 1 s, and 82 nm @ 21 s, as confirmed through a comparative analysis of two independent systems. The advanced long-distance ranging technology is expected to have immediate implications for space research initiatives, such as the space telescope array~\cite{white2000imaging, cash2000laboratory, gendreau2003maxim} and the satellite gravimetry~\cite{flechtner2021satellite, reigber2002high, tapley2004grace}.
\end{abstract}

\section{Introduction}
\begin{multicols}{2}
	Achieving precise long-distance ranging requires a delicate balance between an extended ambiguity range and high precision. Precise ranging relies on accurate timing to determine distance. Interferometry methods, such as continuous-wave (c.w.) laser interferometry, achieve sub-nanometer resolution by measuring the phase change of an optical signal along its path~\cite{nagano2004displacement, pierce2008intersatellite, abich2019orbit}. However, due to the periodicity of the phase, the ambiguity range is limited to half a wavelength. Laser ranging techniques based on pulsed or radio-frequency-modulated signals, as well as the microwave ranging based on GNSS signals, can achieve a larger ambiguity range through time-of-flight measurements~\cite{dickey1994lunar, murphy2013lunar, li2021single, liang20141550, mccarthy2013kilometer,beck2005synthetic, kroes2005precise, allende2016robust}. Nevertheless, because of the constraints imposed by phase and time resolution, their best precision is only in the order of sub-millimeter.
	
	In principle, the synchronization of phases from radio frequencies to optical frequencies enables the achievement of both high precision and a wide ambiguity range. This is crucial for certain applications, such as satellite formation flying, black hole imaging, tests of general relativity, and geographical research~\cite{white2000imaging, cash2000laboratory, gendreau2003maxim, lawson2007terrestrial, turyshev2009laser,pritchard2002satellite, flechtner2021satellite, reigber2002high, tapley2004grace}. For example, to achieve a large synthetic aperture in the X-ray band, the measurement of the absolute baseline distance must be obtained with nanoscale precision~\cite{white2000imaging, cash2000laboratory, gendreau2003maxim}.
	
	Optical frequency combs (OFCs) effectively fulfill this requirement by serving as efficient gears that establish connections between radio frequencies and optical frequencies~\cite{diddams2020optical}. In recent advancements, researchers have explored various ranging methods based on OFCs~\cite{minoshima2000high, ye2004absolute, joo2006absolute, cui2011long, van2012many, na2020ultrafast, lee2010time, lee2013absolute, wang2020long, coddington2009rapid, trocha2018ultrafast, suh2018soliton,caldwell2022time}; however, only a few of them have been demonstrated over long-distance outdoor paths~\cite{lee2010time, wang2020long, lee2013absolute}. The dual-comb ranging method, which leverages the phase interference between adjacent teeth of OFCs, exhibits the potential to extend the ambiguity range beyond 100 km. The phase locking of OFCs with the stable source, such as the ultra-stable laser, is anticipated to achieve accuracy in the tens of nanometers simultaneously, with a fractional ranging resolution of about ${10}^{-13}$. However, due to high transmission loss and noise (atmospheric disturbance), it is extremely challenging to measure such a long distance with the accuracy expected by theory. So far, the reported longest absolute distance measurement based on dual-comb ranging is still less than 1 km~\cite{lee2013absolute}.
\end{multicols}

\section{Two-way dual-comb ranging}
\begin{figure}[H]
	\centering
	\begin{minipage}[b]{1\linewidth}
		\subfigure{
			\includegraphics[width=0.7\columnwidth]{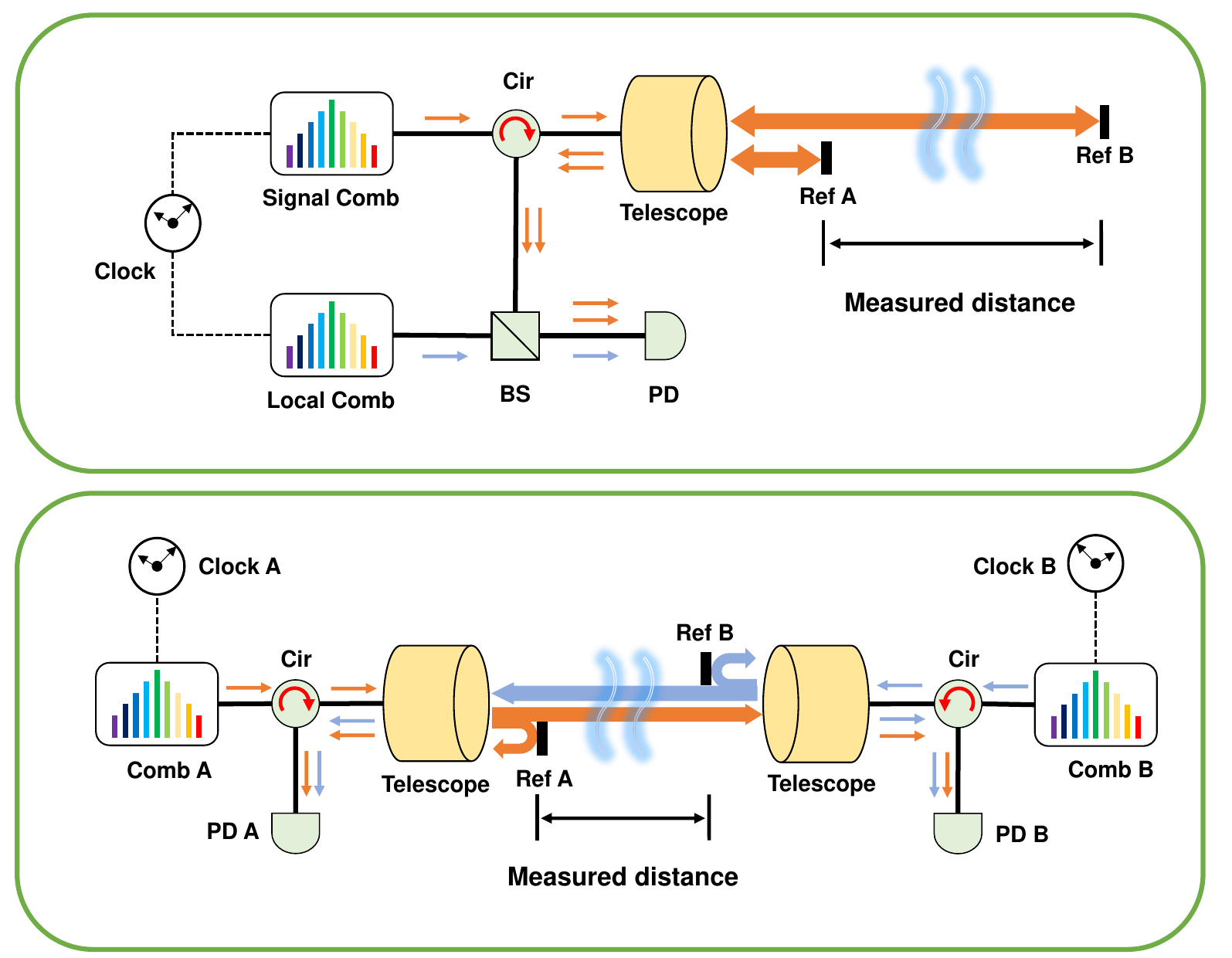}
			\put(-305,222){\small\textbf{A}}
			\put(-305,95){\small\textbf{B}}
		}
		\centering
		\subfigure{
			\includegraphics[width=0.35\columnwidth]{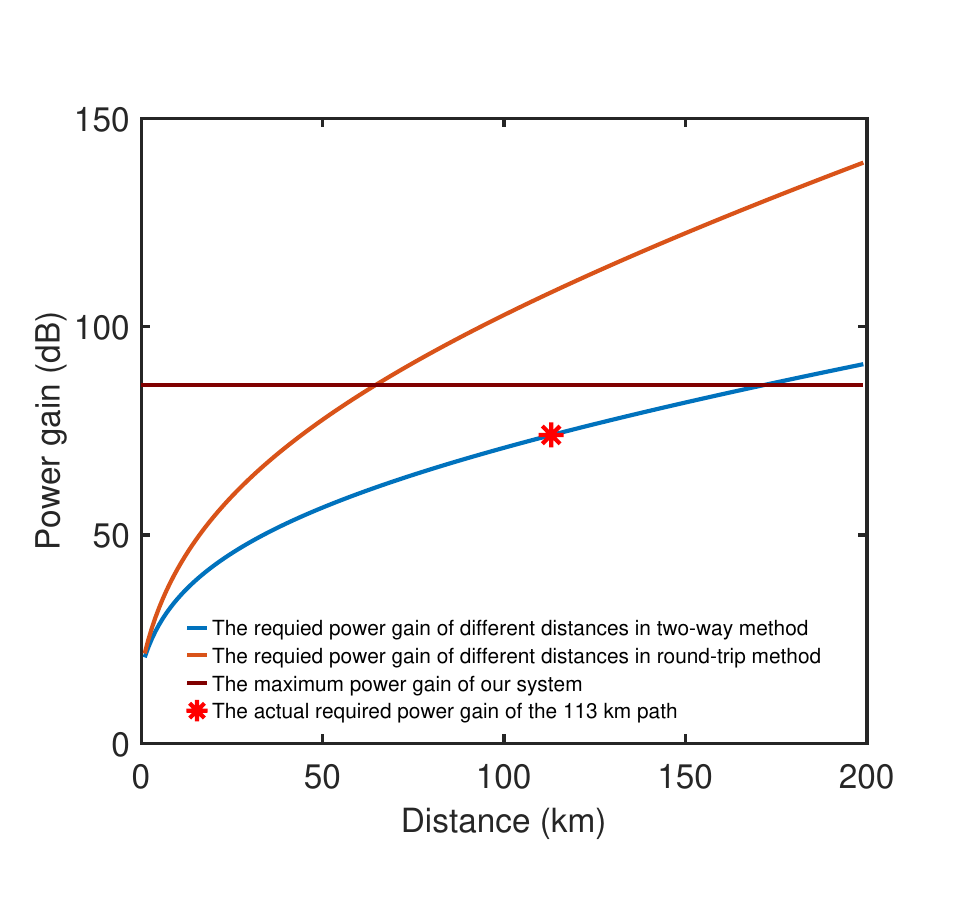}
			\put(-133,113){\small\textbf{C}}
		}
		\subfigure{
			\includegraphics[width=0.35\columnwidth]{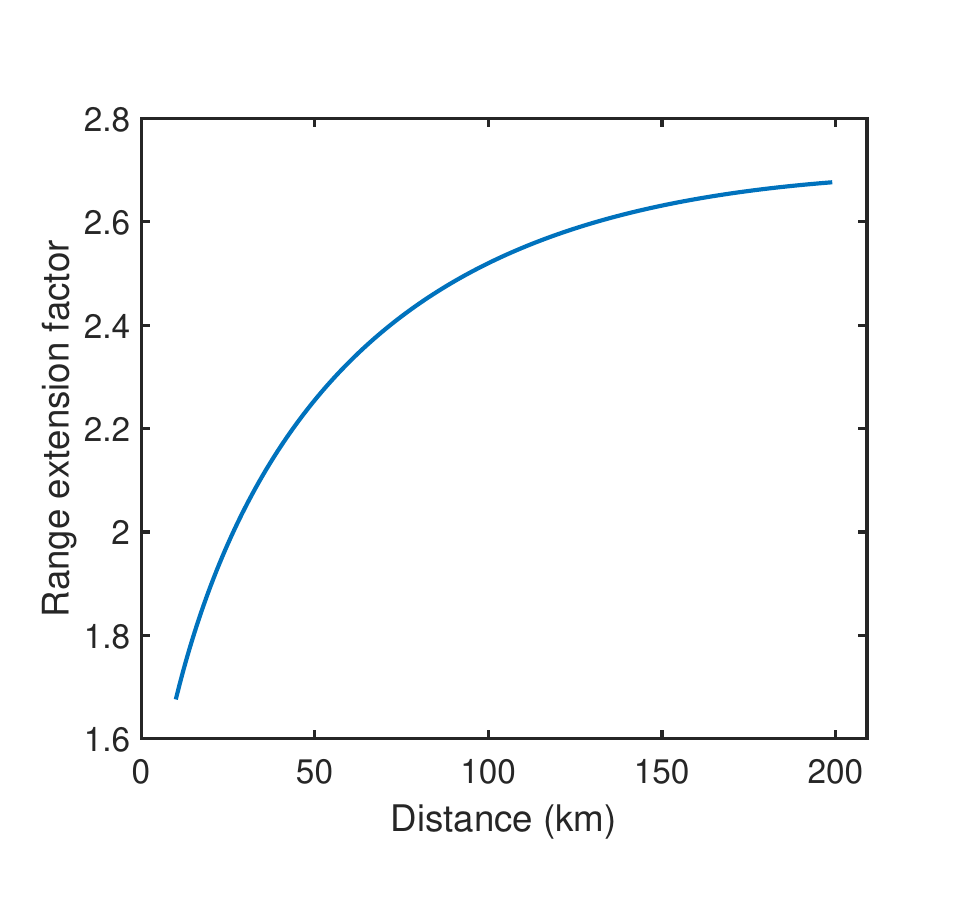}
			\put(-133,113){\small\textbf{D}}
		}
	\end{minipage}
	\caption{\textbf{Two methods of dual-comb absolute ranging.} (A) Round-trip method: The light from signal comb (orange) is reflected by both reference plane A and reference plane B, resulting in interference with the local comb (blue) to obtain distance information. (B) Two-way method: Comb A (orange) and comb B (blue) are individually phase-locked to the local clock. The interference between the light reflected from the local reference plane and the light transmitted from the opposite end is utilized for extracting distance information. (C) The required power gain at different distances. The power gain required for a 113 km distance using the two-way method is equivalent to that needed for a 44 km distance in the round-trip method. (D) The relationship between the range extension factor and the distance (measured by the two-way method). Here, the range extension factor is defined as the ratio between the measured distance obtained using the two-way method and that obtained using the round-trip method.}
	\label{fig1}
\end{figure}

\begin{multicols}{2}
	In this work, we propose a two-way dual-comb ranging (TWDCR) approach to extend the ranging distance by mitigating the effects of transmission noise and loss. The traditional round-trip dual-comb ranging method and TWDCR are shown in Fig.~\ref{fig1}A and Fig.~\ref{fig1}B, respectively. In the round-trip method, the light from signal comb is reflected by both reference plane A and reference plane B, resulting in interference with the local comb to obtain distance information. In the two-way method, comb A and comb B are individually phase-locked to the local clock. The interference between the light reflected from the local reference plane and the light transmitted from the opposite end is utilized for extracting distance information. Compared to the round-trip method, the advantage of the two-way method lies in the fact that the traveling signal only passes the distance once and is not limited by the diameter of the remote reflector (Ref B in Fig.~\ref{fig1}A), thus avoiding the power loss of the signal outside the reflector~\cite{han2024dual}. The power gains required at different distances for both methods are depicted in Fig.~\ref{fig1}C. In our experiment, the actual required power gain for a 113 km path is 74 dB~\cite{shen2022free} (See Supplementary Materials for details). As shown in Fig.~\ref{fig1}D, for distances exceeding 100 km, the two-way approach can extend the range by at least 2.5 times with identical signals and detection sensitivity. This two-way method increases complexity of the setup and data extraction process due to the additional timing combinations on both ends and the need for prior time alignment using time-frequency transfer~\cite{shen2022free}. However, this trade-off can be acceptable for certain applications that are primarily influenced by power loss. For instance, in the satellite gravimetry, the detection of Earth's gravitational field can be achieved through inter-satellite ranging between GEO and LEO satellites~\cite{reigber2002high}. Based on the parameters of our current experimental setup, the power loss of the round-trip method is 106 dB, while the power loss of the two-way method is 53 dB (See Supplementary Materials for details). So in this application, the power loss requirement of the round-trip method is difficult to meet, and the two-way method is suitable.
	
	In the TWDCR, terminals A and B are equipped with OFCs that synchronize to stable clocks. These two OFCs have a slight different repetition rate (a few kilohertz) to ensure regular interference. At each terminal, a telescope is employed for receiving and transmitting OFC signals. A fraction of the launched OFC signal is reflected by a reference mirror, serving as the local reference, while the other part traverses the distance to be measured and is collected by the telescope at the other terminal. Interference occurs between the local reference and the traveling signal at each terminal. The distance to be measured refers to the length between two reference mirrors. The timing results extracted from the interferograms at terminals A and B are expressed as ${T}_{A} = T_{L} + {\tau}_{BA}$ and ${T}_{B} = T_{L} - {\tau}_{BA}$, where \(T_{L}\) represents the flight time of light between two reference mirrors, and \({\tau}_{BA}\) denotes the time difference between clocks at terminals A and B. The distance can be determined from the timing data, the velocity of light \(c\), and the refractive index of the medium \(n\). Subsequently, the distance is calculated as $L = c({T}_{A} + {T}_{B})/{2n}$.
\end{multicols}

\section{Setup}
\begin{figure}[H]
	\centering
	\includegraphics[width=0.88\linewidth]{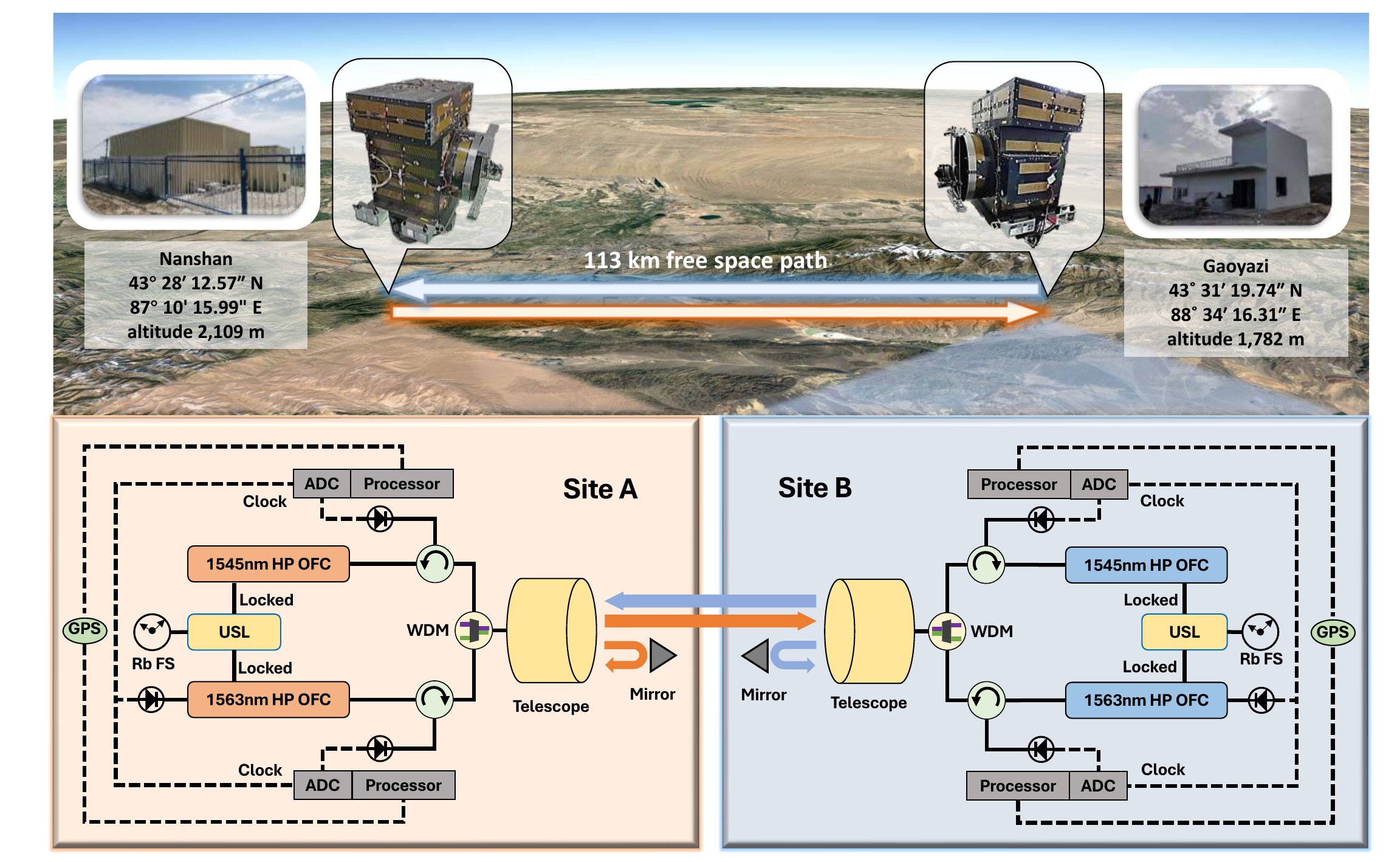}
	\put(-397,232){\textbf{A}}
	\put(-397,116){\textbf{B}}
	\caption{\textbf{The experimental setup.} (A) Overview of the 113 km experimental path for TWDCR. (B) The primary apparatus of two experimental sites. The OFCs at two terminals of the same wavelength have a slight frequency difference, forming a set of equipment for TWDCR. The performance of the TWDCR method is evaluated by comparing the results of two independent ranging devices at different wavelengths.  Some abbreviations are: USL, Ultra-Stable Laser; HP OFC, High-Power Optical Frequency Comb; WDM, Wavelength Division Multiplex; GPS, Global Positioning System; ADC, Analog-to-Digital Converter; Rb FS, Rubidium Frequency Standard.}
	\label{fig2}
\end{figure}

\begin{multicols}{2}
	We conducted a ranging experiment at a distance of 113 km to validate the TWDCR approach in Urumqi, Xinjiang Province. The performance of the TWDCR method is evaluated by comparing the results of two independent ranging devices at different wavelengths. Fig.~\ref{fig2}A illustrates the location information, while Fig.~\ref{fig2}B depicts the experimental setup. Both terminals are equipped with identical instruments, including a 1550.12 nm ultra-stable laser with a stability of 3E-15 @ 1 s, two high-power (1 Watt) OFCs with a spectral span of about 10 nm and wavelengths centered at 1545 nm and 1563 nm respectively, a GPS receiver, a rubidium atomic clock with an accuracy of 5E-11, and a telescope with an aperture diameter of 400 mm. In each terminal, both OFCs with different wavelengths are phase-locked to the ultra-stable laser. The carrier-envelope offset (CEO) frequency and the beat frequency of OFCs, referenced to the atomic clock, are set equal to cancel each other out (See Methods for details). In this experimental design, the repetition frequency of OFCs is uniquely determined by the frequency of the ultra-stable laser at each terminal. The frequency difference of the ultra-stable lasers located at both terminals was initially controlled within tens of hertz using time-frequency dissemination~\cite{shen2022free}. The relative drift of the frequency difference is about 0.1 Hz/s, which has a negligible influence on ranging accuracy during data acquisition time. The repetition frequencies of the 1545 nm OFCs are set at 249.999 973 MHz (${f}_{r,1545}$) and 249.999 973 MHz + 2.585 kHz (${f}_{r,1545}+\Delta{f}_{r,1545}$), while the repetition frequencies of the 1563 nm OFCs are set at 199.995 018 MHz (${f}_{r,1563}$) and 199.995 018 MHz + 2.068 kHz (${f}_{r,1563}+\Delta{f}_{r,1563}$).

	The OFC signals of different wavelengths are combined and separated at each terminal by using a wavelength-division multiplexer (WDM). A common-mode path is established between the WDMs for the two TWDCR systems operating at different wavelengths. In this path, a corner reflector with a diameter of 2 inches is positioned at a distance of 0.1 meters from the telescope to serve as the reference for ranging at each terminal. The OFC signal reflected by the corner reflector and that transmitted from the opposite terminal are wavelength-dependently guided by the WDM at each terminal and finally coupled into the corresponding pigtail collimators.
	
	In addition to using the TWDCR method, high-power OFCs, and large-aperture telescopes, we also utilize low-noise photodetectors (PDs) and accurate data acquisition and processing systems to overcome the challenges of high power loss.  The low-noise PDs receive approximately 20 $\mu$W of OFC signal reflected from the local corner reflector and approximately 40 nW of OFC signal from the opposite terminal. The received power of 40 nW aligns well with the estimated path loss. The equivalent noise power of the PD is 3.5 pW$/\sqrt{Hz}$, enabling a high detection sensitivity below 3 nW. The interferograms are digitized by a 14-bit analog-to-digital converter and recorded by using a field-programmable gate array (FPGA). Timing data is extracted using linear optical sampling (LOS)~\cite{coddington2009rapid}, which includes Fast Fourier Transform (FFT), phase decoupling and extraction of interference waveforms (See Methods for details). The optical distance is obtained based on the timing data of the two sites. The start time of data collection at the two sites is within 30 ns by using the Global Positioning System (GPS). Electronic devices at each site are synchronized with the repetition frequency of the 1563 nm comb, serving as a common frequency reference.
\end{multicols}

\begin{multicols}{2}
	Another challenge lies in achieving a substantial ambiguity range in the presence of high noise caused by air turbulence and atmospheric changes along the 113 km free space path. The Time Deviation (TDEV) of the time-of flight over open-air path is measured at approximately 69 fs @ 1s and 3.5 ps @ 300s, while the time jitter ranges between 40-60 fs~\cite{shen2022free}. By comparing the results of two independent OFC interference systems, we can effectively capture real-time atmospheric drift and mitigate its impact. The ambiguity range depends on the period of the OFC signal through the path~\cite{coddington2009rapid}. In our experiments, the OFC signals at both sites passing through the path have a small difference in repetition frequency, resulting in two ambiguity ranges. The values of these parameters at both terminals are approximately 0.6 m for OFCs at 1545 nm, denoted by ${D}_{r1}$ and ${D}_{r2}$, while for the OFCs at 1563 nm, the values of the same parameters are about 0.75 m. They can be determined as follows: ${D}_{r1}=c/2n{f}_{r}$, ${D}_{r2}=c/2n({f}_{r}+\Delta{f}_{r})$. To extend the ambiguity range, we employ the 'synthetic repetition rate' technique and conduct air dispersion analysis.
	
	The technique of 'synthetic repetition rate' is similar to that of 'synthetic wavelength'~\cite{coddington2009rapid,lee2013absolute}. By utilizing two combinations of repetition frequencies, ${f}_{r}$ and ${f}_{r}+\Delta{f}_{r}$, as well as ${f}_{r}$ and ${f}_{r}-\Delta{f}_{r}$ for identical wavelength OFCs at both terminals, and exchanging the two repetition frequencies in each combination between two terminals, we are able to achieve an extended ambiguity range denoted as ${D}_{AR}=({D}_{r1}{D}_{r2}) /2({D}_{r2}-{D}_{r1})$. The ranging distance $L$ within the interval (0, ${D}_{AR}$) can be determined using the formula ${L}={N}_{1}{D}_{r1}/2+{d}_{1}={N}_{2}{D}_{r2}/2+{d}_{2}$, where $N_{1}$ and $N_{2}$ represent the period numbers, ${d}_{1}$ and $d_{2}$ denote distances obtained from the interference waveforms within the ranges of (0,${D}_{r1}/2$) and (0,${D}_{r2}/2$). On the other hand, the exchange in each combination effectively cancels additional ranging asymmetry caused by non-common fiber paths at a sub-meter level (See Methods for details). 
	Based on the broad wavelength feature of OFCs and the air refractive index model~\cite{ciddor1996refractive}, we use the air dispersion analysis to achieve a preliminary ranging result from the phase information of the interferogram, with an ambiguity range of about ${10}^{8}$ km and a precision within 2 km (See Methods for details). The air dispersion analysis provides a rough ranging value of 113 km +/- 2 km.
\end{multicols}

\section{Results}

\begin{figure}[H]
\centering
\begin{minipage}[b]{0.48\linewidth}
	\subfigure{
		\includegraphics[width=1.05\columnwidth]{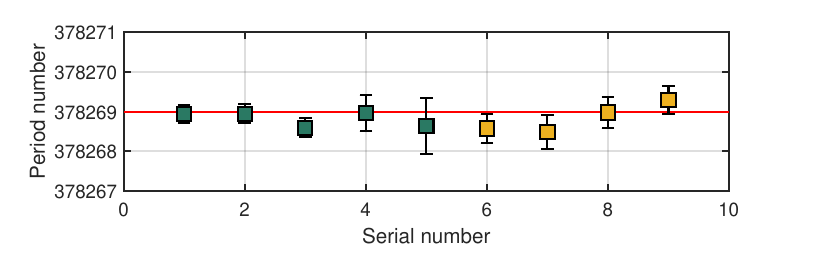}
		\put(-194,67){\small\textbf{A}}
	}
	\subfigure{
		\includegraphics[width=1.05\columnwidth]{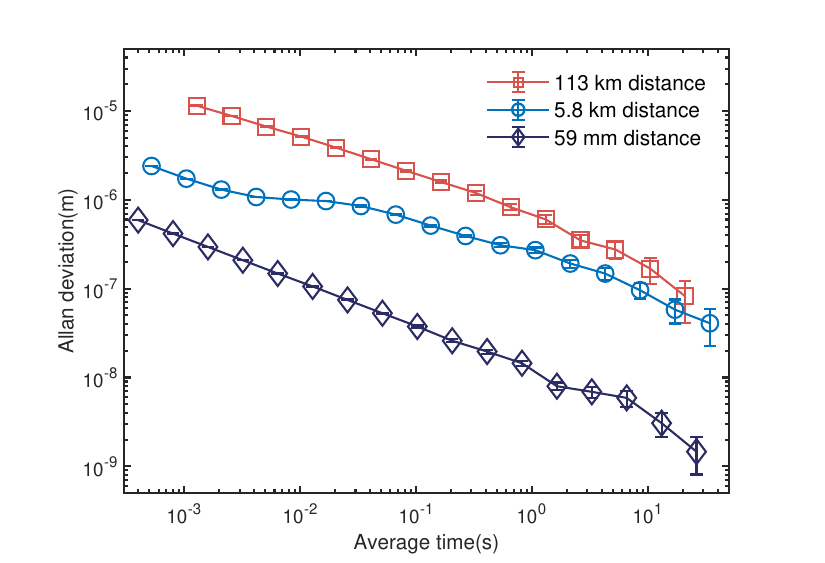}
		\put(-194,146){\small\textbf{B}}
	}
	\hspace{250pt}
\end{minipage}
\begin{minipage}[b]{0.5\linewidth}
	\subfigure{
		\includegraphics[width=0.9\columnwidth]{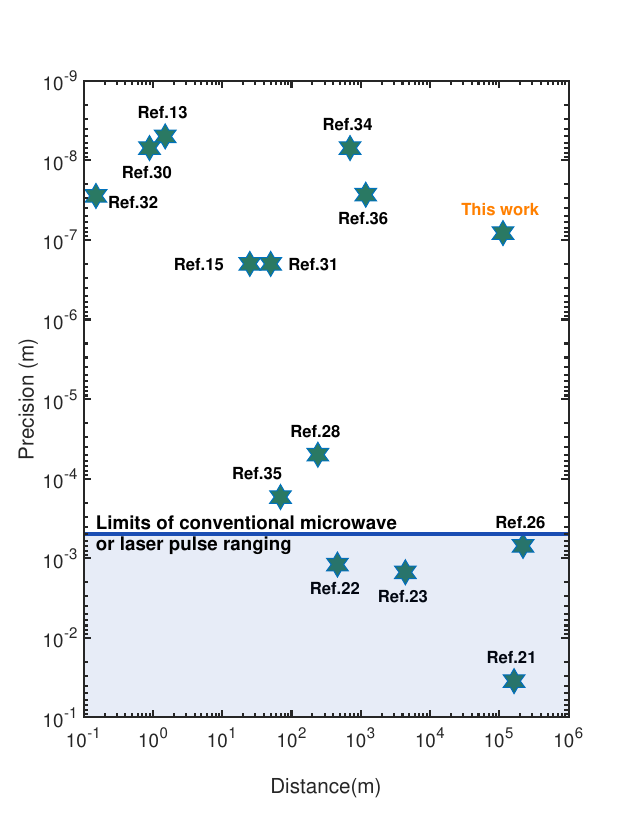}
		\put(-175,245){\small\textbf{C}}
	}
\end{minipage}
\caption{\textbf{Experimental results of TWDCR.} (A) Period number $N_{1}$ of absolute ranging. The absolute distance $L$ is defined as: ${L}={N}_{1}{D}_{r1}/2+{d}_{1}={N}_{2}{D}_{r2}/2+{d}_{2}$, where ${D}_{r1}$ and ${D}_{r2}$ are determined by the parameters of OFCs, ${d}_{1}$ and ${d}_{2}$ are obtained from the interference waveforms within the range of (0,${D}_{r1}/2$) and (0,${D}_{r2}/2$), ${N}_{2}$ is equal to ${N}_{1}+4$, as defined by the preliminary ranging result. Each period (${D}_{r1}/2$) or (${D}_{r2}/2$) corresponds to an approximate distance of 0.3 meters. The data was collected over two consecutive days between 8 PM and 10 PM, with each day represented by the same color. (B) Precision of absolute ranging. The red open square, blue open circle, and purple diamond lines represent the results over paths of 113 km, 5.8 km, and 59 mm, respectively. (C) Results of absolute distance and precision measurements using various methods. The results in the blue area are based on techniques for modulating microwave and laser pulse ranging, while other results are obtained using various OFC methods.} 
%Ref.12 and Ref.14 report the outcomes of microwave-based inter-satellite ranging. Other results indicate the actual distance and precision of absolute free space ranging based on various optical frequency comb schemes.}
\label{fig3}
\end{figure}

\begin{multicols}{2}
	The key to obtaining an absolute distance lies in acquiring integer period values $N_{1}$ and $N_{2}$ through the use of  the technique of synthetic repetition rate. The correctness of the $N_{1}$ and $N_{2}$ is guaranteed by ensuring that each data point reaches the accuracy of $({D}_{r2}-{D}_{r1})/4\sqrt{2}$ in synthesis. The extended ambiguity range and the desired accuracy can be adjusted by varying $\Delta{f}_{r}$. In our specific case, the extended ambiguity range of the 1545 nm OFC system is limited to about 60 km due to the influence of residual atmospheric noise on the measurement accuracy. Here, we have set ${D}_{AR}$ as 30 km, which is shorter than 60 km but longer than the resolution required for air dispersion analysis. The coarse distance value 113 km obtained from the air dispersion analysis indicates that $N_{2}$ is equal to $N_{1}+4$. By utilizing the equation ${L}={N}_{1}{D}_{r1}/2+{d}_{1}={N}_{2}{D}_{r2}/2+{d}_{2}$ and employing two sets of experimentally obtained $\{d_{1}\}$ and $\{d_{2}\}$, the period number $N_{1}=(4D_{r2}+d_{2}-d_{1})/(D_{r1}/2-D_{r2}/2)$ is accurately determined as illustrated in Fig.~\ref{fig3}A. The average value observed is 378,268.82 with a standard deviation of 0.26. It should be noted that all period values $N_{1}$ are consistently around 378269 as the data was collected over consecutive days between 8 PM and 10 PM. Table 1 presents four frequency groups for 1545 nm OFCs in a single measurement of synthetic repetition rate. To obtain sufficient accuracy, the acquisition time for each frequency group is set at 4 minutes. The change between frequency groups takes about 3 minutes, and the total operation time for each measurement is half an hour.	
\end{multicols}

\begin{table}[H]
	\centering	
	\label{tab1}
	\begin{small}
		\begin{tabular}{c|c|c} % l、c、r分别为左对齐、居中对齐、右对齐
			\hline % 添加水平线
			Group number & Parameters & Values \\ % 第一行标题
			\hline % 再次添加水平线
			\multirow{2}{*}{Group 1} & ${f}_{r,1545,A}$ & 249.999 973 MHz\\ % 每一行数据
			& ${fr}_{1545,B}$ & 249.999 973 MHz+2585.3Hz\\
			\hline
			\multirow{2}{*}{Group 2} & ${f}_{r,1545,A}$ & 249.999 973 MHz+2585.3Hz \\
			& ${fr}_{1545,B}$ & 249.999 973 MHz \\
			\hline
			\multirow{2}{*}{Group 3} & ${f}_{r,1545,A}$ & 249.999 973 MHz \\
			& ${fr}_{1545,B}$ & 249.999 973 MHz-2585.3Hz \\
			\hline
			\multirow{2}{*}{Group 4} & ${f}_{r,1545,A}$ & 249.999 973 MHz-2585.3Hz \\
			& ${fr}_{1545,B}$ & 249.999 973 MHz \\
			\hline % 最后一行也需要添加水平线
		\end{tabular}
	\end{small}
	\label{tab1}
	\caption{Frequency for 1545 nm OFCs in a single synthetic repetition rate measurement}
\end{table}
	
\begin{multicols}{2}	
	The ranging precision achieved by comparing the results of two independent OFC interferences is shown in Fig.~\ref{fig3}B. The Allan Deviation (ADEV) for 113 km ranging is 11.5 $\mu$m @ 1.3 ms, 681 nm @ 1 s, and 82 nm @ 21 s. In order to test the system, we also conducted ranging experiments over distances of 69 mm and 5.8 km using the same equipment. The precision for the outdoor path of 5.8 km achieves 274 nm @ 1 s and 41 nm @ 34 s. The system floor, obtained in the 69 mm ranging experiments, is 12 nm @ 1 s and 1.5 nm @26 s. Precision decreases with increasing distance due to the gradual accumulation of atmospheric noise along the measurement path. Additionally, data rates drop with distance, leading to larger errors. 
	
	The experimental results of various ranging methods are presented in Fig.~\ref{fig3}C. The TWDCR approach demonstrates the capability to achieve absolute ranging on a hundred-kilometer scale with sub-micron precision. Compared with other methods, TWDCR significantly improves the absolute distance measurement accuracy by more than several orders of magnitude at the equivalent measurement length. This breakthrough not only meets the requirements of numerous applications, but also has the potential to enhance the performance of certain applications beyond their existing plans. For instance, the accurate positioning of the formation flying satellites in the Micro-Arcsecond X-ray Imaging Mission (MAXIM) requires absolute ranging with sub-micrometer accuracy~\cite{white2000imaging,cash2000laboratory,gendreau2003maxim}. The length of the inter-satellite baseline directly impacts the angular resolution of the telescope. The 1 km baseline corresponds to an angular resolution of about ${10}^{-7}$ arcseconds, which is two orders of magnitude better than the finest resolution images ever achieved (in the radio part of the spectrum) and over one million times better than the current best X-ray images. Our approach is expected to increase the length of the baseline beyond 100 km, resulting in an angular resolution of about ${10}^{-9}$ arcseconds, which is two orders of magnitude higher than initially idea. Furthermore, accurate inter-satellite absolute ranging enables precise satellite positioning and orbit tracking, which is a significant addition to satellite gravimetry~\cite{flechtner2021satellite, reigber2002high,tapley2004grace}. Compared to existing interferometry methods, our approach offers the advantage of avoiding cycle slips in inter-satellite ranging results. This allows us to instantaneously capture the complete variations of the Earth's gravity field during large-scale changes in geology, such as earthquakes, floods, and volcanic eruptions.
\end{multicols}

\begin{figure}[H]
	\centering
	\includegraphics[width=0.6\linewidth]{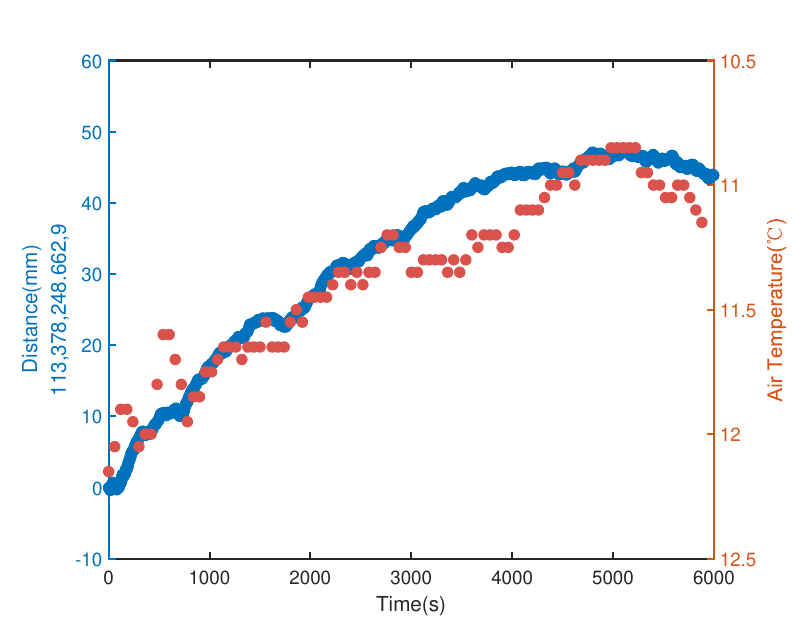}
	\caption{\textbf{Long-term ranging results of TWDCR.} By combining the measured values ${d}_{1}$ at each moment, the calculated values ${N}_{1}$ and the average meteorological data, the absolute distance values $L$ can be obtained. The distance results (blue) show an initial value of 113,378,248.662,9 mm with an overall fluctuation of 50 mm. The average temperatures (red) obtained from the two weather stations also exhibit a similar trend with the observed variation in distance.}
	\label{fig4}
\end{figure}

\begin{multicols}{2}
By combining the measured values ${d}_{1}$ at each moment, the calculated values ${N}_{1}$ and the meteorological data~\cite{ciddor1996refractive}, the absolute distance values $L$ can be obtained. Fig.~\ref{fig4} shows the 6000 s absolute ranging results of our systems. We monitor meteorological conditions with two weather stations (PC-8) located at each terminal and determine the absolute distance using the average meteorological data. As previously mentioned, ranging is based on measuring the time of flight, so its accuracy is limited by uncertainties in measuring the air refractive index at a level of approximately ${10}^{-7}$. In principle, the influence of air refractive index can potentially be mitigated by deploying additional weather monitoring stations along the path. Then, the accuracy of absolute ranging may be constrained by the atmospheric refractive index model derived from meteorological data at about ${10}^{-8}$. Further improvements can be achieved through the use of dual-comb spectroscopy~\cite{han2024dual}, which accurately determines certain atmospheric conditions, or two-color method~\cite{wu2013extremely} that does not rely on environmental parameters. Fortunately, the space telescope array and satellite gravimetry operate in space at least 400 km above Earth's surface, experiencing an extremely low pressure level of ${10}^{-8}$ Pa. The uncertainty associated with air reflective index in these space applications is below ${10}^{-16}$, thereby enabling our systems to attain a fractional ranging uncertainty of ${7.3}\times {10}^{-13}$.
\end{multicols}

\section{Conclusion}
\begin{multicols}{2}
	In summary, we propose a novel TWDCR method to extend the ranging distance by mitigating the effects of transmission noise and loss. Compared to traditional round-trip dual-comb ranging method, this innovative approach significantly reduces power loss, enabling a 2.5-fold extension in measuring distance. In addition, we used high-power OFCs, large-aperture telescopes, low-noise photodetectors (PDs) and precise data acquisition and processing systems in our experiments, thereby achieving precise absolute ranging distances exceeding 100 km. To overcome transmission noise, we employ a combination of air dispersion analysis technique and synthetic repetition rate method. Air dispersion analysis offers an ambiguity range of ${10}^{8}$ km with a resolution of 2 km, while synthetic repetition rate method provides nanometer precision along with an ambiguity range of 30 km. By using these techniques, we have demonstrated high-precision absolute distance measurements over a path of 113 km with a precision of 82 nm @ 21 s. To the best of our knowledge, this study represents the first instance of such precise absolute distance measurement over a path exceeding 100 km, providing essential technical foundations for future inter-satellite absolute distance measurements. This technology is expected to improve the angular resolution of space telescope arrays and prevent the cycle slips of satellite gravimetry.
\end{multicols}

\clearpage

\section*{Acknowledgements}
%\begin{multicols}{2}
This research was supported by the National Natural Science Foundation of China (grant no. T2125010, 61825505); National Key Research and Development Programme of China (grant no. 2020YFA0309800, 2020YFC2200103); Strategic Priority Research Programme of Chinese Academy of Sciences (grant no. XDB35030000); Anhui Initiative in Quantum Information Technologies (grant no. AHY010100); Key R$\&$D Plan of Shandong Province (grant no. 2021ZDPT01); Shanghai Municipal Science and Technology Major Project (grant no. 2019SHZDZX01); Innovation Programme for Quantum Science and Technology (grant no. 2021ZD0300100); Tian-shan Talent training Program (grant no. 2023TSYCLJ0053); Fundamental Research Funds for the Central Universities; and the New Cornerstone Science Foundation through the XPLORER PRIZE.
%\end{multicols}

\clearpage
\section*{Methods and Supplementary Materials}
\setcounter{equation}{0}%将公式编号归0
\renewcommand{\theequation}{S.\arabic{equation}}
\renewcommand{\thetable}{S.\arabic{table}}
\setcounter{figure}{0}%将公式编号归0
\renewcommand{\thefigure}{S.\arabic{figure}}

\textbf{Principle of the TWDCR.} 
The linear optical sampling method of TWDCR  enables the accurate acquisition of time and distance information within a single interference period. In linear optical sampling, the electric fields of OFCs in terminals A and B are expressed as:
\begin{equation}
	\label{E:freq1}
	\begin{aligned}
		{E}_{AP} \left(t\right)&=\sum_{k} {E}_{A, k} \exp \left[i2\pi\left(\nu_{A}+k{f}_{A}\right)\left( t - {T}_{CRA} - {T}_{L} - {T}_{DRB}\right)\right]\\ 
		{E}_{AR} \left(t\right)&=\sum_{k} {E}_{A, k} \exp \left[i2\pi\left(\nu_{A}+k{f}_{A}\right)\left( t - {T}_{CRA} - {T}_{DRA}\right)\right]\\ 
		{E}_{BP} \left(t\right)&=\sum_{k} {E}_{B, k} \exp \left[i2\pi\left(\nu_{B}+k{f}_{B}\right)\left( t - {T}_{CRB} - {T}_{L} - {T}_{DRA} -{\tau}_{BA}\right)\right]\\ 
		{E}_{BR} \left(t\right)&=\sum_{k} {E}_{B, k} \exp \left[i2\pi\left(\nu_{B}+k{f}_{B}\right)\left( t - {T}_{CRB} - {T}_{DRB} - {\tau}_{BA}\right)\right]\\ 
	\end{aligned}
\end{equation}
where the electric field ${E}_{AP} \left(t\right)$ represents the portion of the comb A laser that arrives at terminal B, while ${E}_{AR}\left(t\right)$ represents the portion of the comb A laser that is reflected from the local reference in terminal A. Similarly, ${E}_{BP}\left(t\right)$ and ${E}_{BR}\left(t\right)$ represent corresponding components of the comb B laser. The variables $T_{CRA}$, $T_{DRA}$, $T_{CRB}$ and $T_{DRB}$ denote the flight time from the local OFC to the local reference plane and from the local reference plane to the detector at terminal A and terminal B, respectively. The clock difference between terminals A and B is ${\tau}_{BA}$. The flight time of the distance between the two reference planes is defined as ${T}_{L}$. The repetition frequencies $f_{A}$ and $f_{B}$ are approximately equal to $f_{r}$ with a slight difference $\Delta f_{r}$. Interference signals are detected using photodetectors (PDs). The voltage output of PD in terminal A is:
\begin{small}
	\begin{equation}
		\label{Volt1}
		\begin{aligned}
			&{V}_{A}(t)\propto \operatorname{Im} \left[E_{AR}^{*}(t) E_{BP}(t) \right]=\exp\left[i2\pi\left(\nu_{B}-\nu_{A} \right)t\right]\\
			&\times\exp\left[i2\pi\nu_{A}\left({T}_{CRA}+{T}_{DRA}\right)-i2\pi\nu_{B}\left({T}_{CRB}+{T}_{L}+{T}_{DRA}+{\tau}_{BA}\right)\right]\\
			&\times\sum_{k}{E}_{A, k} {E}_{B, k} \exp\left[ i2\pi k \Delta f_{r} t+i2\pi k{f}_{A}\left({T}_{CRA}+{T}_{DRA}\right)-i2\pi k{f}_{B}\left({T}_{CRB}+{T}_{L}+{T}_{DRA}+{\tau}_{BA}\right) \right]
		\end{aligned}
	\end{equation}
\end{small}
while the voltage output of the PD in terminal B is:
\begin{small}
	\begin{equation}
		\label{Volt2}
		\begin{aligned}
			&{V}_{B}(t)\propto \operatorname{Im} \left[E_{AP}^{*}(t) E_{BR}(t) \right]=\exp\left[i2\pi\left(\nu_{B}-\nu_{A}\right)t\right]\\
			&\times\exp\left[i2\pi\nu_{A}\left({T}_{CRA}+{T}_{L}+{T}_{DRB}\right)-i2\pi\nu_{B}\left({T}_{CRB}+{T}_{DRB}+{\tau}_{BA}\right)\right]\\
			&\times\sum_{k}{E}_{A, k} {E}_{B, k} \exp\left[ i2\pi k \Delta f_{r} t+i2\pi k{f}_{A}\left({T}_{CRA}+{T}_{L}+{T}_{DRB}\right)-i2\pi k{f}_{B}\left({T}_{CRB}+{T}_{DRB}+{\tau}_{BA}\right) \right]
		\end{aligned}
	\end{equation}
\end{small}
The time information is extracted with high precision from the interference phase. By using Fast Fourier Transform (FFT), the phase information $\Phi$ of the $k$ comb tooth can be calculated as:
\begin{equation}
	\label{phase1}
	\begin{aligned}
		{\Phi}_{A}\left(k,{T}_{L}\right)&=2\pi\nu_{A}\left({T}_{CRA}+{T}_{DRA}\right)-2\pi\nu_{B}\left({T}_{CRB}+{T}_{L}+{T}_{DRA}+{\tau}_{BA}\right)\\
		&+2\pi k{f}_{A}\left({T}_{CRA}+{T}_{DRA}\right)-2\pi k{f}_{B}\left({T}_{CRB}+{T}_{L}+{T}_{DRA}+{\tau}_{BA}\right) \\
		{\Phi}_{B}\left(k,{T}_{L}\right)&=2\pi\nu_{A}\left({T}_{CRA}+{T}_{L}+{T}_{DRB}\right)-2\pi\nu_{B}\left({T}_{CRB}+{T}_{DRB}+{\tau}_{BA}\right)\\
		&+2\pi k{f}_{A}\left({T}_{CRA}+{T}_{L}+{T}_{DRB}\right)-2\pi k{f}_{B}\left({T}_{CRB}+{T}_{DRB}+{\tau}_{BA}\right)
	\end{aligned}
\end{equation}
By subtracting ${\Phi}_{A}\left(k,{T}_{L}\right)$ from ${\Phi}_{B}\left(k,{T}_{L}\right)$, we get:
\begin{equation}
	\label{phase2}
	\begin{aligned}
		{\Delta\Phi}\left(k,{T}_{L}\right)&=2\pi\left(\nu_{A}-\nu_{B}\right)
		\left({T}_{DRB}-{T}_{DRA}\right)+2\pi\left(\nu_{A}+\nu_{B}\right){T}_{L}\\
		&+2\pi k\left({f}_{A}-{f}_{B}\right)
		\left({T}_{DRB}-{T}_{DRA}\right)+2\pi k\left({f}_{A}+{f}_{B}\right){T}_{L} \\
	\end{aligned}
\end{equation}
The value of ${T}_{L}$ can be calculated in the formula by using the phase change as the comb tooth serial number $k$ increases. However, the presence of a term $2\pi k\left({f}_{A}-{f}_{B}\right)\left({T}_{DRB}-{T}_{DRA}\right)$ in the formula introduces error due to the difference in path lengths between the reference plane and the detector at terminal A and B. To cancel this error, we interchanged the values of the repetition rates ${f}_{A}$, ${f}_{B}$ and merged both sets of data into a unified dataset. The change of the repetition frequency values does not affect the values of ${\nu}_{A}$ and ${\nu}_{B}$. Assuming that ${\nu}_{A}+{\nu}_{B}= 2{\nu}_{0}$, the phase is calculated as follows:
\begin{equation}
	\label{phase3}
	\begin{aligned}
		{\Delta\Phi}\left(k,{T}_{L}\right)&=8\pi\nu_{0}{T}_{L}+4\pi k\left({f}_{A}+{f}_{B}\right){T}_{L} \\
	\end{aligned}
\end{equation}
\\
\textbf{Effects of atmospheric dispersion.} 
The previously discussion on principles assumes a constant refractive index of air. However, in the actual atmospheric path, it is necessary to consider the multi-order relationship between the refractive index of air and optical frequency due to atmospheric dispersion. In fact, the relationship between ${\Delta\Phi}\left(k,{T}_{L}\right)$ and the serial number $k$ of comb teeth is non-linear. The refractive index of air corresponding to $k$ can be expressed as:
\begin{equation}
	\label{index1}
	\begin{aligned}
		{n}\left(k\right)&={n}_{0}+q\left(\upsilon-\upsilon_{0}\right)+p\left(\upsilon-\upsilon_{0}\right)^{2}+\mathrm{O}\left(\upsilon^{3}\right)\\
		&={n}_{0}+qk{f}_{r}+p\left(k{f}_{r}\right)^{2}+\mathrm{O}\left(\upsilon^{3}\right)
	\end{aligned}
\end{equation}
where $n_{0}$ represents the refractive index of air at the frequency ${\nu}_{0}$. $q$ and $p$ represent the first and second order coefficients of refractive index with respect to optical frequency. Based on Eqn.~\eqref{phase3} and Eqn.~\eqref{index1}, the phase $\Delta\Phi\left(k,L\right)$ can be calculated as follows:
\begin{small}
	\begin{equation}
		\label{phase4}
		\begin{aligned}
			{\Delta\Phi}\left(k,L\right)&=\frac{4\pi L} {c}[2{n}_{0}\nu_{0}+\left({n}_{0}{f}_{A}+{n}_{0}{f}_{B}+q{\nu}_{0}{f}_{A}+q{\nu}_{0}{f}_{B}\right)k \\
			&+\left(q+p{\nu}_0\right)\left({{f}_{A}}^{2}+{{f}_{B}}^{2}\right){k}^{2}+p\left({{f}_{A}}^{3}+{{f}_{B}}^{3}\right){k}^{3}] 
		\end{aligned}
	\end{equation}
\end{small}
The resulting phase in linear optical sampling can only be achieved within the interval $(0,2\pi)$. In fact, the range of distance $L$ for each order coefficient of $k$ is also limited by the phase range. We use $d_{0}$, $d_{1}$ and $d_{2}$ instead of $L$ to represent the residues of different ambiguity ranges corresponding to the different order coefficients. Here, the distance interval for the zero-order coefficient is $(0, c/4{n}_{0}{\nu}_{0})$, which corresponds to an optical wavelength-level span. According to the air refractive index model~\cite{ciddor1996refractive}, $q$ is approximately $5.6 \times 10^{-21}$ and $p$ is about $1.5 \times 10^{-35}$ at room temperature. Therefore, the interval of $L$ for first-order coefficient becomes $(0, c/2{n}_{0}({f}_{A}+{f}_{B}))$, corresponding to a distance range at $(0, 0.3 m)$, while the range of second-order coefficient is $(0, c/[2(q+p{\nu}_{0})({f}_{A}^{2}+{f}_{B}^{2})])$, corresponding to a distance range at $(0, 1.4 \times 10^{11} m)$. The coefficients of the third-order terms of k can be disregarded at approximately 100 km due to their negligible magnitude compared to the first and second order terms. Consequently, Eqn.~\eqref{phase4} can be modified as follows:
\begin{small}
	\begin{equation}
		\label{phase5}
		\begin{aligned}
			{\Delta\Phi}\left(k,L\right)&=\frac{4\pi} {c}[2{n}_{0}\nu_{0}{d}_{0}+\left({n}_{0}{f}_{A}+{n}_{0}{f}_{B}+q{\nu}_{0}{f}_{A}+q{\nu}_{0}{f}_{B}\right){d}_{1}k +\left(q+p{\nu}_0\right)\left({{f}_{A}}^{2}+{{f}_{B}}^{2}\right){d}_{2}{k}^{2}] 
		\end{aligned}
	\end{equation}
\end{small}
It is worth noting that the quadratic term of Eqn.~\eqref{phase5} can be ignored when the measured distance falls within the range of millimeters to meters. However, as the absolute distance $L$ extends from the laboratory path to the outdoor kilometer-level path, the ratio of the second-order coefficient to the first-order coefficient increases continuously. Therefore, a quadratic function needs to be used for fitting the relationship between ${\Delta\Phi}\left(k,{T}_{L}\right)$ and $k$. 
\\

\begin{figure}[H]
	\centering
	\includegraphics[width=0.88\linewidth]{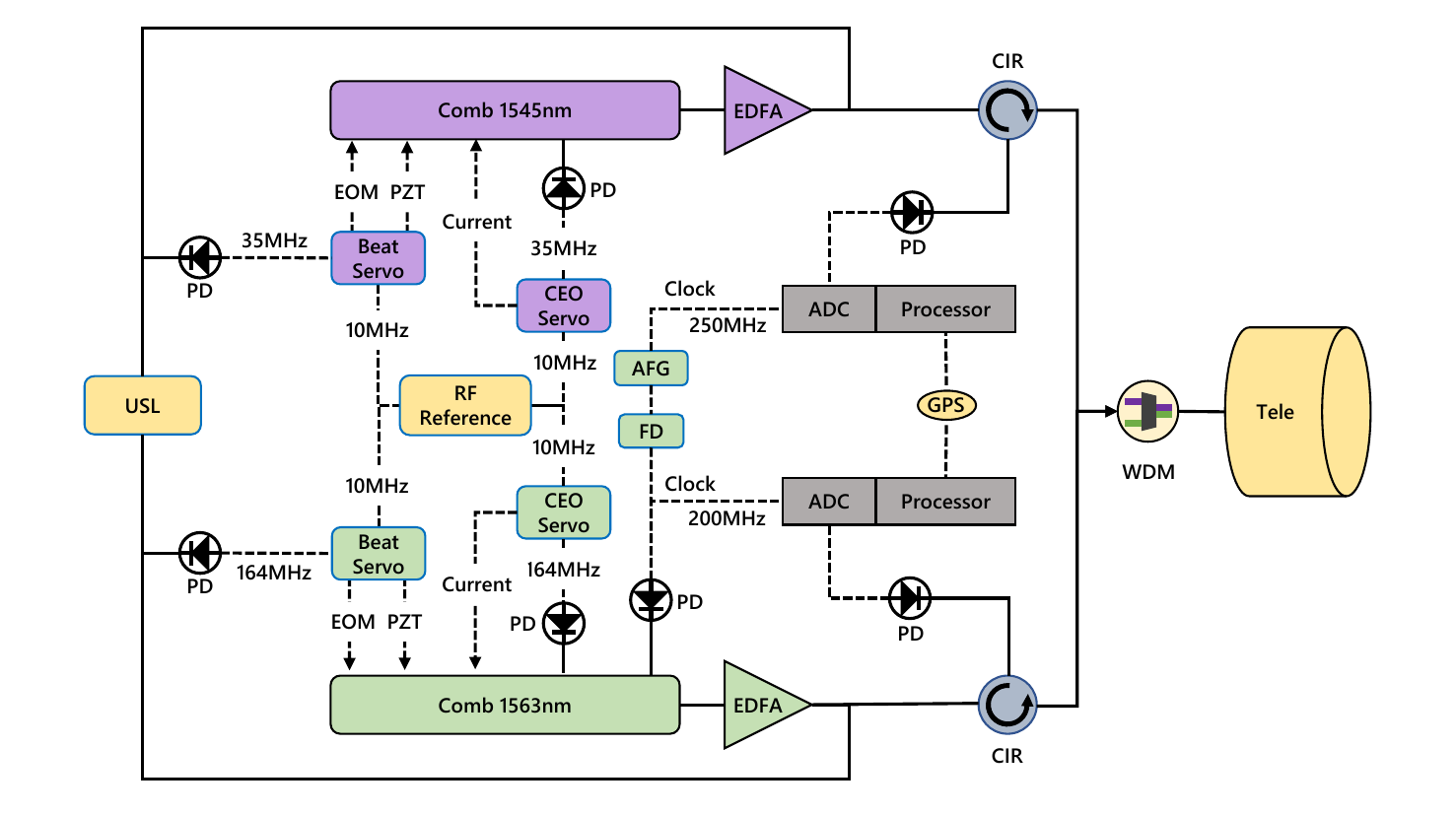}
	\caption{\textbf{Detailed experimental setup of single terminal.} The solid lines represent the optical signals, while the dashed lines represent the electronic signals. Some abbreviations are: USL, Ultra-stable laser; EDFA, Erbium-doped fiber amplifier; WDM, Wavelength Division Multiplex; GPS, Global Positioning System; ADC, Analog-to-Digital Converter; PD, Photodetector; CIR, Circulator; Tele, Telescope; FD, Frequency Divider; AFG, Arbitrary Function Generator. }
	\label{fig5}
\end{figure}

\textbf{Details of our experimental setup.} Fig.~\ref{fig5} shows the experimental setup of one terminal. The setups of both terminals are nearly identical. Each terminal comprises two phase-locked OFCs and one USL. The upper part of the figure illustrates the 1545 nm OFC, while the bottom part represents the 1563 nm OFC. Both OFCs use high-power EDFAs to amplify their power to approximately 1 Watt. The amplified signal is divided into the signal synchronized with USL and the signal utilized for ranging purposes. Circulators are utilized to distinguish between transmitted and received signals. Most local fiber paths are contained within the integrated module that maintains a temperature stability with a standard deviation of about 10 mK. The acquisition rate of electronic devices are synchronized with USL thourgh the repetition rate of the 1563 nm OFC.
\\

\textbf{Power loss of experimental path.} 
The power loss of the entire transmission path, from the output of EDFA at one terminal to the PD at the other terminal can be calculated by using:
\begin{equation}
	\label{loss1}
	\begin{aligned}
		{\eta}&={\eta}_{fiber}{\eta}_{tele}{T}_{atm}{\eta}_{sm}\left({\frac{{D}_{r}}{L\theta}}\right)^{2}\\
	\end{aligned}
\end{equation}
where ${\eta}_{fiber}$ represents the loss of local fiber paths at both terminals, which is approximately 4 dB. The loss of two telescopes, denoted as ${\eta}_{tele}$, is around 7.8 dB. Atmospheric transmittance ${T}_{atm}$~\cite{strohbehn1978laser}, accounting for air absorption and scattering of the propagating beam along our path, is approximately 12.5 dB. ${\eta}_{sm}$ represents the single-mode fiber coupling efficiency for free-space optical path through atmospheric turbulence~\cite{dikmelik2005fiber,andrews2005laser},which amounts to 27.9 dB in our systems. The geometric attenuation, expressed as $\left({D}_{r}/{L\theta}\right)^{2}$, is 21.8 dB. The typical attenuation value for the entire path is 74 dB, but it may vary between a lower limit of 66 dB or an upper limit of 83 dB depending on weather conditions.
\\

\textbf{Power loss in inter-satellite ranging between GEO and LEO satellites.} 
Here we calculate the geometrical loss between the GEO and LEO satellites using both the two-way method and round-trip method, respectively. In two-way method, the geometrical loss can be represented as follows:
\begin{equation}
	\label{loss1}
	\begin{aligned}
		{\eta}_{tw}&=\left({\frac{D_{t}}{L\theta}}\right)^{2}\\
	\end{aligned}
\end{equation}
where $D_{t}$ repersents the aperture of telescope, $L$ is the inter-satellite distance between GEO and LEO satellites, $\theta$ represent the effective transmitter full-angle divergence. The values of parameters could be estimated as $D_{t}$ = 400 mm, $L$ = 36000 km, $\theta$ = 5 ${\mu}$rad, and the loss is calculated to be 53 dB. In round-trip method, the geometrical loss can be represented as follows:
\begin{equation}
	\label{loss1}
	\begin{aligned}
		{\eta}_{rt}&=\left({\frac{D_{t}}{L\theta}}\right)^{2}·\left({\frac{D_{m}}{L\theta}}\right)^{2}\\
	\end{aligned}
\end{equation}
The aperture of reflector $D_{m}$ can be estimated to be the same as that of telescope $D_{t}$, and the loss is calculated to be 106 dB.
	
\clearpage

\bibliographystyle{unsrt}
\bibliography{distance}

%\begin{thebibliography}{10}
%	\expandafter\ifx\csname url\endcsname\relax
%	\def\url#1{\texttt{#1}}\fi
%	\expandafter\ifx\csname urlprefix\endcsname\relax\def\urlprefix{URL }\fi
%	\expandafter\ifx\csname href\endcsname\relax
%	\def\href#1#2{#2} \def\path#1{#1}\fi
%\end{thebibliography}

%\bibliographystyle{elsarticle-num}
%\bibliography{DCS}

\end{document}